\documentclass[draft,grl]{agu2001}
\usepackage{graphicx,balu}
\def\cite#1{[\citeauthor{#1}]}
\authorrunninghead{NADIGA}
\titlerunninghead{Oceanic Zonal Jets}
\authoraddr{Balasubramanya T. Nadiga, MS-B296, Los Alamos, NM-87545; balu@lanl.gov}

\begin{document}

\setkeys{Gin}{draft=false}
\title{On Zonal Jets in Oceans}
\author{Balasubramanya T. Nadiga}
\affil{Los Alamos National Laboratory, MS-B296, Los Alamos, NM-87545}
\begin{abstract}

  We find that in parameter regimes relevant to the recently observed
  alternating zonal jets in oceans, the formation of these jets can be
  explained as due to an arrest of the turbulent inverse-cascade of
  energy by {\em free} Rossby waves (as opposed to Rossby {\em basin} modes)
  and a subsequent redirection of that energy into zonal modes. This
  mechanism, originally studied in the context of alternating jets in
  Jovian atmospheres and two dimensional turbulence in
  zonally-periodic configurations survives in spite of the presence of
  the meridional boundaries in the oceanic context.
\end{abstract}

%
%

%

\begin{article}

\section{Introduction}
A proposed explanation of the alternating zonal jets in Jovian
atmospheres is that they are due to a tendency
of turbulence in thin shells on the surface of a rotating sphere to
organize itself into zonal jets (e.g., see \cite{vasavada, galperin}).
The anisotropic jets result from an interplay between an inverse
cascade of energy [\citeauthor{kraichnan, charney}] and the
latitudinal variation of the vertical component of planetary rotation
(e.g., see [\citeauthor{newell,rhines}]). While baroclinic instability
and convective processes are thought to be the main sources of
small scale energy, classical geostrophic turbulence theory
[\citeauthor{charney}] predicts a cascade of this energy (vertically)
to larger scales as well in a process that has been termed
barotropization. Hence, in the context of this explanation of
atmospheric-zonal jets, they have been simulated and
studied extensively using the barotropic vorticity equation on either
the doubly-periodic or zonally-periodic beta-plane or on the surface
of a sphere using forced-dissipative settings. 
In these settings, the effect of geometry on dynamics is minimized
in the sense that the flow in the zonal direction, the direction in
which the dynamics of the Rossby waves are highly asymmetric, is
homogeneous.  Dynamically, the formation of the zonal jets in this
homogenous setting is thought to involve certain kinds of resonant
interactions (sideband triad and quartet) of Rossby waves packets
whose amplitudes are slowly varying functions of space and time
[\citeauthor{newell}].

More recently, observational \cite{maxi} and computational
\cite{nakano} evidence point to the occurence of multiple alternating
zonal jets in the world oceans as well. However, the dynamics
underlying their formation is not clear.

On the one hand, given that the governing equations are the same in
the atmospheric and oceanic contexts, it would not be unreasonable to
expect, from a turbulence point of view, that the same dynamical
mechanism---Rossby wave dispersion arresting the inverse cascade of
energy and redirecting it into zonal modes---underlies the phenomenon,
be it in the ocean or in the atmosphere.  Clearly, unlike the constant
stratification of the atmosphere, surface-intensified stratification
in the oceans inhibits full barotropization (e.g., see \cite{flierl}).
Nevertheless, the importance of the barotropic mode (with a
thermocline depth of 1 km in a 5 km deep ocean) is clearly borne out
in Fig. 2 and Table 1 of Fu \& Flierl [1980] and other such studies
confirm an inverse cascade of barotropic kinetic energy.  High
vertical coherence of jet structure in models \cite{nakano,maxi}
further suggests the importance of barotropic dynamics.

On the other hand, the presence of boundaries can, besides being able
to support viscous boundary layers and act as sources/sinks of
enstrophy, allow for new (inviscid) mechanisms. For example, in a
closed basin, (a) Fofonoff gyres arise as statistical equilibrium
solutions of the barotropic vorticity equation, and (b) Rossby basin
modes arise, resonant interactions of which have been studied as
mechanisms for generating both mean flows (starting with Pedlosky
[1965]) and mesoscale variability (starting with [Harrison and Robinson
[1979]). Such mechanisms could possiblly generate alternating zonal
jets as well.  Interestingly, LaCasce [2002] finds that the arrest of
the inverse cascade of energy by basin normal modes is largely
isotropic.  However, in a recent article studying rectification
processes in a three layer quasi-geostropic beta plane basin, Berloff
[2005] concludes that the alternating zonal jets he found in that
setting were most likely driven by nonlinear interactions between some
meridionally structured baroclinic basin modes and some secondary
(i.e., related to finite amplitude background flows) basin modes. If
this were to be the most important mechanism for the formation of
alternating zonal jets in ocean basins, by involving
spatially-nonlocal (basin) modes this mechanism would be fundamentally
different from the (spatially) local arguments of turbulence that are
usually thought to apply in the atmospheric context.



In this letter, we demonstrate that in parameter regimes relevant to
alternating zonal jets in the oceans, such jets can be formed by {\em
  free} Rossby waves (as opposed to Rossby {\em basin} modes)
arresting the inverse-cascade of energy. We then go on to show that
the jet width scales well with Rhines' scale.  This suggests that the
dynamics of alternating zonal jets in oceans are likely local and in
this sense similar to those in previously studied atmospheric
contexts.  We suggest that the nonlocal
resonant-interaction-of-basin-modes mechanism becomes more important
at larger values of turbulent kinetic energy (TKE).  Curiously, only
the latter regime has been investigated before within the framework of
the barotropic vorticity equation \cite{lacasce}, and as far as we
know this is the first time that alternating zonal jets have been
obtained in a closed basin using the barotropic vorticity equation.


The rest of the letter is structured as follows: The next section
briefly describes the modeling approach, and the following one
presents computational results. As a matter of convenience, and with
no loss of generality, these two sections consider the governing
equations and present the results in a nondimensional form. The final
section establishes the correspondence between the nondimensional
parameter values considered and their dimensional counterparts in
actual ocean settings.

\section{Modeling Approach} 
We consider the barotropic vorticity equation
\begin{equation}
{\partial q \over \partial t} + J({\psi}, q) = {\it F} + {\it D}
\label{bve}
\end{equation}
for the evolution of barotropic potential vorticity $q = \zeta +
{\beta}y = {\nabla^2}{\psi} + {\beta}y$, where $\zeta$ is
relative vorticity, ${\psi}$ is velocity streamfunction, ${\it F}$
is forcing, ${\it D}$ is dissipation and $J(~,~)$ is the Jacobian
operator given by $J({\psi}, q) = -{\partial {\psi} \over \partial y}
{\partial q \over
  \partial x} + {\partial {\psi} \over \partial x} {\partial q \over
  \partial y}$.  The above equation is considered on a midlatitude
beta plane with a latitudinal gradient of the vertical component of
rotation of $\beta$; y-coordinate increases northwards and
x-coordinate eastwards in a closed square basin, $2\pi$ on a side,
discretized into 1024{\sc x}1024 cells. An energy and enstrophy
conserving finite-differencing is used with Runge-Kutta timestepping
[\citeauthor{GN}].
Given the inverse-cascade of energy of 2D turbulence, forcing $F$ is
concentrated around a high wavenumber $k_f$, as a combination of sines
and cosines consistent with the boundary conditions used.  Their
amplitudes $\sigma$, drawn randomly from a Gaussian distribution, are
delta-correlated in time resulting in $F={\sigma(t)}/{\sqrt{\delta t}}
f(k_f,t)$ with an energy input rate $\epsilon$ of
$\sigma^2\int\!\!\int f \nabla^{-2}f dx dy$ and an enstropy input rate
$\eta$ of $k_f^2 \epsilon$. In all the computations presented, given
the domain size of $2\pi${\sc x}$2\pi$ discretized into 1024{\sc x}1024
cells, 
$k_{max}$ is 512 and
$k_f$ is between 128 and 129.

Dissipation $D = -\nu_p \nabla^{2p} \zeta -\nu_0 \zeta$, consisting of
a small-scale-selective component to dissipate the (largely)
downscale-cascading enstrophy input at the forcing scale, and Rayleigh
friction component that mainly acts to dissipate the (largely)
upscale-cascading energy.  At lateral boundaries, besides no
through-flow, we use superslip boundary conditions.
The coefficient $\nu_p$ is
diagnosed dynamically in terms of the enstrophy input rate as
$\nu_p=C_K\eta^\frac{1}{3}\Delta x^{2p}$, using Kolmogorov-like ideas
and a Kolmogorov scale of $\Delta x$.


Given the above setup, the problem consists of three important
parameters: $\beta$, $\epsilon$ and $\nu_0$. We briefly recall a few
relevant spatial scales in terms of these parameters.  First, in
purely two-dimensional turbulence, a Kolmogorov scale for the
dissipation of energy may be obtained using the usual arguments as
\bea 
k_{fr}=(3 C_K)^\frac{3}{2}\left(\frac{\nu_0^3}{\epsilon}\right)^\frac{1}{2} 
\approx 50 \left(\frac{\nu_0^3}{\epsilon}\right)^\frac{1}{2} 
\eea 
(e.g., see \cite{danilov}). In the absence of $\beta$ this would be
the scale at which Rayleigh friction would act to stop the inverse
cascade of energy.  However, in the presence of $\beta$, Rossby wave
dispersion can instead arrest the inverse cascade and
redirect energy into zonal modes. In the absence of large scale
friction, and under the assumption that the spectral flux of energy in
the inverse-cascade inertial range is determined by the energy input
rate $\epsilon$ (due to forcing), this would happen at
$k_\beta=(\beta^3/\epsilon)^{1/5}$ [\citeauthor{maltrud}].
If, however, energy is concentrated near $k_\beta$, this arrest
mechanism would occur at the Rhines' scale
$k^R_\beta=\sqrt{\beta/U_{rms}}$ \cite{rhines} (also obtained by equating
the turbulence frequency $U|k_\beta|$ and the Rossby wave frequency
$-\beta \cos\phi/|k_\beta|$, where $\phi=\tan^{-1}{k_y}/{k_x}$).
Given the largely upscale-cascading nature of energy, the
small-scale-selective dissipation operator plays a relatively minor
role in dissipating energy, so that $dE/dt\approx\epsilon-2\nu_0E$,
with energy leveling off at about $\epsilon/2\nu_0$.  Using this
energy balance in the expression for Rhines' scale leads to
\cite{danilov,shafer} \bea
k^R_\beta=\left(\frac{\beta}{2}\right)^\frac{1}{2}
\left(\frac{\nu_0}{\epsilon}\right)^\frac{1}{4}.  \eea

\section{Results}
\begin{figure}
\noindent\includegraphics[width=20pc]{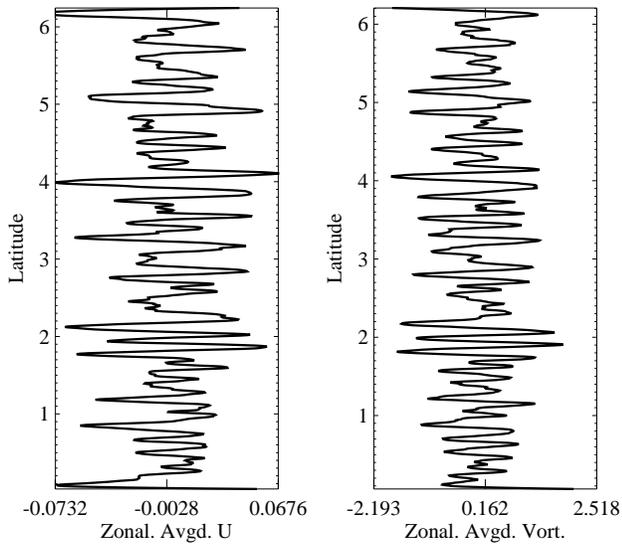}
\caption{\label{bb21d} Meridional plot of the instantaneous zonally-averaged zonal-velocity and vorticity.}
\end{figure}

\begin{figure} 
\noindent{\includegraphics[width=20pc]{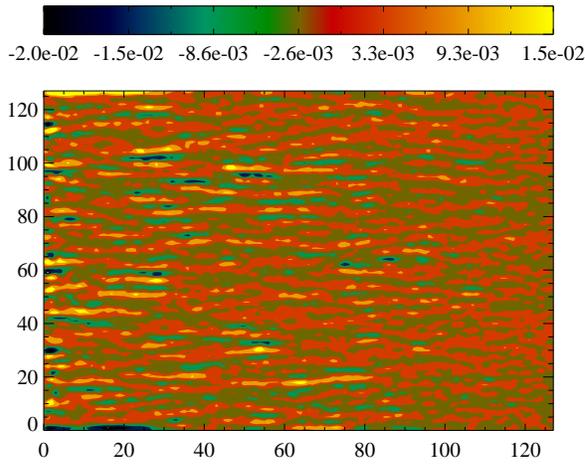}}
\caption{\label{bb21a} The alternating zonal jets are evident in the
  time-averaged, two-dimensional zonal-velocity field. While forcing is
  homogeneous, jets are more prominent in western  regions.}
\end{figure}

\begin{figure}
\noindent{\includegraphics[width=20pc]{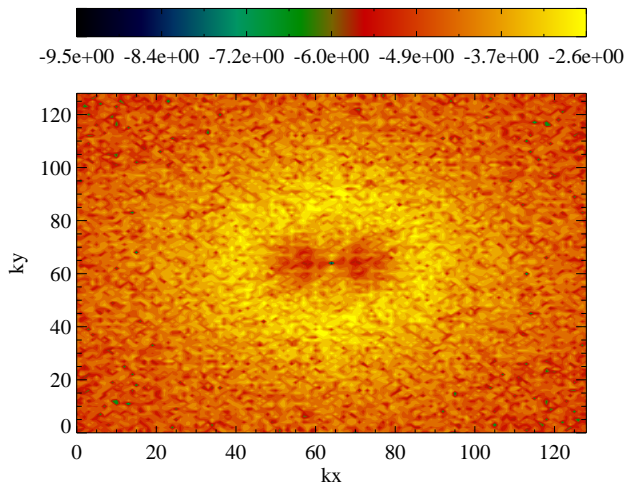}}
\caption{\label{bb21g} The time-averaged two-dimensional energy spectrum displays the familiar anisotropic 'dumbell' shape.}
\end{figure}

Table 1 gives the basic parameters and the derived scales discussed
above for a series of simulations.  In all the cases considered, care
is taken to ensure that the spectrum of the zonal component of energy
has equilibrated. While there are important differences between some
of the cases in Table 1, we postpone a detailed discussion of these
differences to a later article, and go on to examine a representative
case---case C presently. An examination of the instantaneous,
zonally-averaged, zonal-velocity and relative-vorticity fields plotted
as a function of latitude in Fig.~\ref{bb21d} suggests alternating
zonal jets of a characteristic width.  To further verify this, we
examine a few other familiar diagnostics.  First, Fig.~\ref{bb21a}
shows the time-mean two-dimensional zonal-velocity field
after the flow has reached statistically-stationarity,
and the alternating zonal jets are evident in this figure.  
Note that a) even though the forcing is homogeneous, the jets are more
pronounced to the west, b) unlike with observations, time-mean jet signatures are
obtained and analysed, and c) the geometry of the jets
are not significantly different when the time-varying components are
analysed (not shown).
Finally, while the
meridional gradient of time-averaged potential-vorticity is dominated
by $\beta$ (stable), the instantaneous flow quite frequently violates
the barotropic stability criterion $u_{yy}<\beta$.
That these alternating zonal jets are related to anisotropization of
the inverse cascade of energy of two dimensional turbulence by Rossby
wave dispersion is verified by the the dumbell shape near the
origin, characteristic of the process (e.g., see
\cite{maltrud}), in the contour plot of the two dimensional
spectral density of energy in Fig.~\ref{bb21g}.

\begin{figure}
\noindent\includegraphics[width=20pc]{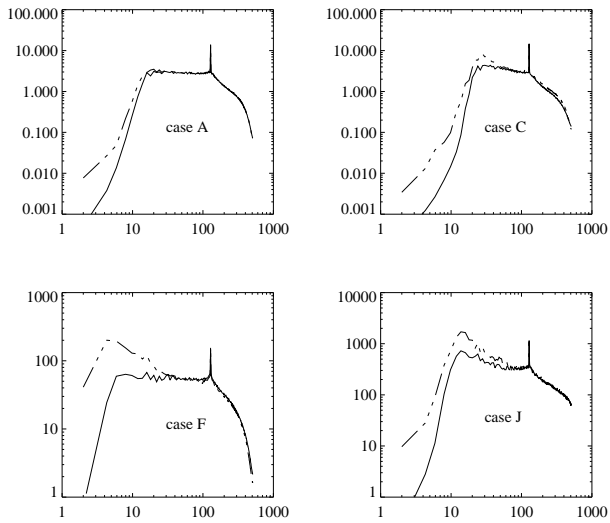}
\caption{\label{bb-spec} Time-averaged one-dimensional zonal 
  (dot-dashed line) and residual (solid line) energy spectra. Both have a
  $k^{-5/3}$ compensation. See text for details.}
\end{figure}

It is not our intent to verify various universal scalings of spectra
in this problem, but to use it as a diagnostic to further confirm the
nature of the dynamics. To this end, we show in Fig.~\ref{bb-spec} the
range of spectra that we obtain in the parameter range considered.
These figures show the 1D energy spectra averaged over an angle of
${\pi}/{6}$ around $\phi=0$ (residual flow) and $\phi={\pi}/{2}$
(zonal flows) \cite{chekhlov}.  Both the residual and zonal spectra
have futher been compensated for the $k^{-5/3}$ scaling. (A
compensation for $\epsilon^{2/3}$---following Kolmogorov scaling
$E(k)=C_k\epsilon^{2/3}k^{-5/3}$---is avoided since while that would
be appropriate for the residual component, it would not be appropriate
for a possibly different scaling of the zonal component such as
$E_z(k) = C_z \beta^2 k^{-5}$. However, $\epsilon^{2/3}$ compensation has been
applied to the residual spectra to establish the value of the
Kolmogorov constant $C_k$).  A common and important feature of all the
cases is that in the inverse-cascade regime, while at the
high-wavenumber end, the zonal and residual spectra scale similarly,
at lower wavenumbers
the zonal spectra lie above the residual spectra and show 
steepening before they peak. This behavior is as expected and verified
by various investigators in the periodic case relevant to the
atmosphere.  Furthermore, like in the computations of Danilov and
Gurarie [2001], our spectra display significant non-universal
behavior. For example, while in cases A and F, the residual spectra
clearly verify the classic Kolmogorov scaling $C_k
\epsilon^{2/3}k^{-5/3}$ with a Kolmogorov constant $C_k$ of about 6,
that is not the case for cases C and J.  We note parenthetically that
(a) the distribution of spectral energy flux as a function of
wavenumber (not shown) bears remarkable resemblance to that derived
using Aviso, TOPEX/Poseidon, and ERS-1/2 data \cite{scott}, and (b) on
using the spectral flux of energy as a function of wavenumber, as
opposed to a constant value, $C_k$ remains close to 6 in the high
wavenumber range of the inverse-cascade, but then begins to rise at
the lower wavenumbers.

\begin{figure}
\noindent{\includegraphics[width=20pc]{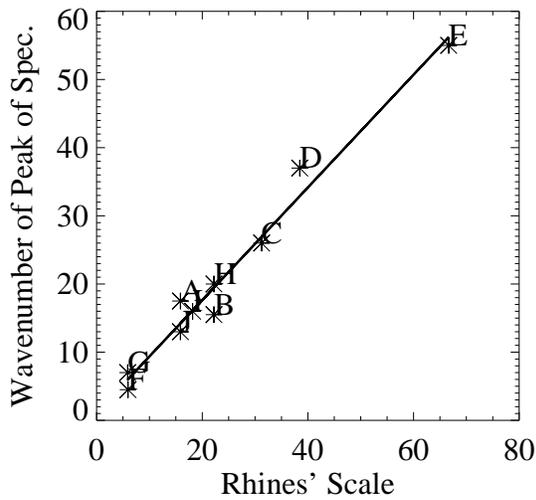}}
\caption{\label{kp-kr} A plot of the wavenumber at which the 
  zonal-spectrum peaks, $k_p$ against the Rhines' scale $k^R_\beta$.
  Symbols correspond to the cases in Table 1 and line to the linear
  least squares fit.}
\end{figure}

Next, we identify the jet width with the wavenumber at which the
(uncompensated) zonal-spectrum peaks, $k_p$, and verify it by
referring to physical-space pictures like in Figs.~\ref{bb21d} and
\ref{bb21a}.  This number is recorded for each of the cases in the
last column of Table~1.  We note, that the wavenumber at which the
residual spectrum peaks is close to this wavenumber as well. In
Fig.~\ref{kp-kr}, we plot the above measure of jet width ($k_p$)
against the Rhines' scale ($k^R_\beta$) and find excellent agreement,
like in the periodic (atmospheric) case (e.g., see Vallis and Maltrud [1993] 
and Danilov and Gurarie [2001].

\section{Discussion}

The simulations and analyses presented in the previous section show
clearly that there are parameter regimes wherein the dynamics of the
alternating zonal jets in a midlatitude ocean basin are not controlled
in a fundamental manner by meridional boundaries. That is to say, in
these parameter regimes, the dynamics of the jets are governed largely
by spatially local interactions, and the arrest of the inverse-cascade
is mediated by free Rossby waves as opposed to Rossby basin modes.
However, we still need to check if such a parameter regime is of
relevance to the oceans in order to establish the importance of this
mechanism in the oceans.

The pronounced signature of the observed jets in western regions of
ocean basins (Fig. 1 of \cite{maxi}) would generally be attributed to
the elevated levels of TKE in the separated western boundary current
(WBC) regions. However, our simulations use homogeneous forcing but
still display similar enhanced jet signatures in the west.  This leads
us to suspect that the enhanced signature of the jets in the west is
more due to its internal dynamics, rather than due to the elevated
levels of TKE in the WBC regions, and that the jets are controlled
more by the ambient (lower) levels of {\em interior} TKE.  An approximate
range of 25 to 100 cm$^2$/s$^2$ is obtained for the latter by
examining an altimetry-derived TKE map for the North Atlantic [pers.
comm.; Scott, 2006].  Keeping this in mind, first consider case C
discussed extensively above: the peak wavenumber $k_p$ is 26 (Table
1); using the observed \cite{maxi} dominant wavelength of 280 km.
leads to $L_{ref}$ of 1160 km. Then, using a typical midlatitude value
of $\beta_{ref}$ of 2 10$^{-11}$ m$^{-1}$s$^{-1}$ and an r.m.s. value
of 7.5 cm/s (mid-range) for the domain-averaged {\em interior}
geostrophic velocity anomaly (TKE), leads to a $\beta_{nd}$
(=$\beta_{ref}L_{ref}^2/U_{ref}$) of 360, corresponding well with 320
used for case C.  As for the ranges of parameters considererd, $5\le
k_p \le 55$ (Table 1); using a range of wavelengths of 500 to 250 km
and domain-averaged {\em interior} TKE level corresponding to $5\le
U_{rms}\le 10$ cm/s, gives $\beta_{nd}$ in the range 30--1900 (c.f.
range of 80--1280 in Table 1). In light of this, we suggest that the
local mechanism wherein the arrest of the inverse cascade is mediated
by free Rossby waves may be important in explaining the formation of
alternating zonal jets in the world oceans.  Obviously, further work
is necessary to {\em definitively} establish the relevance of this mechanism
to the oceans.  The author thanks Boris Galperin for discussions and
the referees for their criticism.

\end{article}\newpage
\begin{table}
\caption{Basic parameters, derived scales and the jet-width wavenumber for the
ten simulations considered}
\begin{flushleft}
\begin{tabular}{c|c|c|c|c|c|c|c}
\tableline
Case&$\beta$&$\epsilon$&$\nu_0$&$k_\beta$&$k_{fr}$&$k_\beta^R$&$k_p$ \\
\tableline
\tableline
A & 80 & 0.50 & 0.1 & 15.9 & 2.2 & 15.9 & 17.5\\
B & 160 & 0.50 & 0.1 & 24.1 & 2.2 & 22.2 & 15.5\\
C & 320 & 0.50 & 0.1 & 36.6 & 2.2 & 31.3 & 26\\
D & 640 & 0.50 & 0.1 & 55.5 & 2.2 & 38.5& 37\\
E & 1280 & 0.50 & 0.1 & 84.0 & 2.2 & 66.7 & 55\\
F & 80 & 33.5 & 0.1 & 6.0 & 0.2 & 5.96 & 4.5\\
G & 80 & 128 & 0.4 & 4.6 & 0.8 & 5.88 & 7\\
H & 1280 & 126 & 0.4 & 27.8 & 1.1 & 22.2 & 20\\
I & 1280 & 296 & 0.4 & 23.4 & 0.7 & 18.2 & 16\\
J & 1280 & 513 & 0.4 & 21.0 & 0.6 & 15.9 & 13\\
\tableline
\multicolumn{1}{c}{\phantom{JJJJJ}} & \multicolumn{1}{c}{\phantom{JJJJJ}} &
\multicolumn{1}{c}{\phantom{JJJJJ}} & \multicolumn{1}{c}{\phantom{JJJJJ}} &
\multicolumn{1}{c}{\phantom{JJJJJ}} & \multicolumn{1}{c}{\phantom{JJJJJ}} &
\multicolumn{1}{c}{\phantom{JJJJJ}} & \multicolumn{1}{c}{\phantom{JJJJJ}} \\
\end{tabular}
\end{flushleft}
\end{table}
%
%

%
%


\end{document}